\documentclass[12pt,authoryear,final]{elsarticle}
\usepackage{longtable}
\usepackage{booktabs}
\usepackage[latin2]{inputenc}
\usepackage[T1]{fontenc}
\usepackage{graphicx}
\usepackage{ae}
\usepackage{aecompl}
\usepackage{setspace}
\usepackage{amsmath}
\usepackage{amssymb}
\usepackage{amsthm}
\usepackage{natbib}
\usepackage{todonotes}
\usepackage{parskip}
    \makeatletter
    \def\ps@pprintTitle{%
      \let\@oddhead\@empty
      \let\@evenhead\@empty
      \def\@oddfoot{\reset@font\hfil\thepage\hfil}
      \let\@evenfoot\@oddfoot
    }
    \makeatother

\title{Identifying Combinatorial Biomarkers by Association Rule Mining in the CAMD Alzheimer's Database}
\author[p]{Balázs Szalkai}
\ead{szalkai@pitgroup.org}
\author[s]{Vince K. Grolmusz}
\ead{grolmusz@enzim.hu}
\author[p,u]{Vince I. Grolmusz\corref{cor1}}
\ead{grolmusz@pitgroup.org}
\cortext[cor1]{Corresponding author}
\author{Coalition Against Major Diseases\corref{cor2}}
\cortext[cor2]{Data used in the preparation of this article were obtained from the Coalition Against Major Diseases database (CAMD). As such, the investigators within CAMD contributed to the design and implementation of the CAMD database and/or provided data, but did not participate in the analysis of the data or the writing of this report.}
 
\address[p]{Protein Information Technology Group, Eötvös University, H-1117 Budapest, Hungary}
\address[u]{Uratim Ltd., H-1118 Budapest, Hungary}
\address[s]{2nd Department of Internal Medicine, Semmelweis University, Budapest, Hungary.}

\begin{document}

\begin{abstract}
{\bf Background:} The concept of  combinatorial biomarkers was conceived around 2010: it was noticed that simple biomarkers are often inadequate for recognizing and characterizing complex diseases. 

\noindent{\bf Methods:} Here we present an algorithmic search method for complex biomarkers which may predict or indicate Alzheimer's disease (AD) and other kinds of dementia. 
We applied  data mining techniques that are capable to uncover implication-like logical schemes with detailed quality scoring. Our program SCARF is capable of finding multi-factor relevant association rules automatically. The new SCARF program was applied for the Tucson, Arizona based Critical Path Institute's CAMD database, containing  laboratory and cognitive test data for more than 6000 patients from the placebo arm of clinical trials of large pharmaceutical companies, and consequently, the data is much more reliable than numerous other databases for dementia.

\noindent{\bf Results:} The results suggest connections between liver enzyme-, B12 vitamin~-, sodium- and cholesterol levels and dementia, and also some hematologic parameter-levels and dementia.
\bigskip

\end{abstract}

\maketitle

\section{Introduction}

Dementia is a major problem presently of high-income countries and also an increasing concern of low-income nations worldwide. Though sporadic before the age of 60, its occurrence is doubled by every five years of age thereafter \citep{Bermejo-Pareja2008,Carlo2002}. About 40 percent of the population over 90 are affected, and up to 20 percent of those between 75 and 84 suffer from this condition \citep{Wortmann2012,war2009}.  The most common cause of dementia is Alzheimer's disease (AD). The earliest symptoms of AD include memory problems; disorientation in time and space; and difficulty with calculation, language, concentration and judgment. As the disease evolves, patients may develop severe behavioral abnormalities and may even become psychotic. In the final stages of the disease the sufferers are incapable of self-care and become bed-bound, for years or even decades. 

The diagnosis of AD in the great majority of the cases is done by clinical criteria, using standardized questionnaires \citep{Mossello2010}.  Generally accepted evidences show that neuropathological damage begins more than 20 years before those clinical signs \citep{Jack2009}, and by the time it is diagnosed, a large part of the neurons are already irreversibly lost. 

In the last years, by the combination of cerebrospinal fluid analysis, clinical signs and neuroimaging techniques a quite reliable diagnostic method emerged \citep{Dubois2007}. The method, however, is prohibitively expensive, is not an early warning-type biomarker, and does not seem to be applicable for wide-scale screening of the senior population. 

Very recently, using the combination of usual clinical laboratory data, cognitive impairment questionnaires and blood-based proteomics assays was reported to reliably diagnose AD, without neuroimaging or cerebrospinal fluid assays \citep{O'Bryant2010, O'Bryant2011}. However, early warning biomarkers are still need to be found.

The final goal of ours is finding new combinatorial biomarkers for Alzheimer's disease. In this paper we report our results that may be used to reach this final goal; but presently we are able to show only that certain sets of laboratory data may make dementia (and not AD) more probable, and certain other sets may make dementia less probable.

There are several large databases on Alzheimer's disease available for researchers. The quality of their data obviously depends on the methodology of the research that produced the database in question. The most well-organized, strictly overseen and rigorously documented experiments are perhaps conducted by the order of large pharmaceutical companies in hospitals and clinics in phase 1, 2 and 3 drug trials. Unfortunately, the detailed results of those trials are seldom published (especially those corresponding to unsuccessful drug trials) since they are owned by the companies that ordered the trials. 

In their Alzheimer's disease database the Tucson, Arizona based Critical Path Institute made available the results of the placebo arm of numerous multi-million dollar clinical trials conducted by the order of large pharmacological companies \citep{Romero2009, Romero2010, Rogers2012}. The data of the placebo line of the trials does not contain proprietary information concerning the effects of the novel drugs under trial, but it does contain reliable, well-organized laboratory and cognitive test-data, presumably in much higher quality than other, larger, but perhaps less strictly conducted and controlled studies for AD. 

Data used in the preparation of this article have been obtained from the Coalition Against Major Diseases (CAMD) database \citep{Romero2009}. In 2008, Critical Path Institute, in collaboration with the Engelberg Center for Health Care Reform at the Brookings Institution, formed the Coalition Against Major Diseases (CAMD). The Coalition brings together patient groups, biopharmaceutical companies, and scientists from academia, the US Food and Drug Administration (FDA), the European Medicines Agency (EMA), the National Institute of Neurological Disorders and Stroke (NINDS), and the National Institute on Aging (NIA). Coalition Against Major Diseases (CAMD) includes over 200 scientists from member and non-member organizations. The data available in the CAMD database have been volunteered by both CAMD member companies and non-member organizations.

In contrast with more conservative statistical methods, we applied data mining techniques for data analysis and combinatorial biomarker search. Data mining, as defined in \citep{PDM}, is the analysis of large observational sets of data for finding new, still unsuspected relations with novel, usually high-throughput methods. Frequently, data mining uses large data sets originally collected for uses other than the data mining analysis \citep{PDM}. 

Simple biomarkers (e.g., the high level of glucose in diabetes) show a physiological condition, related to the appearance or the status of a disease. The concept of combinatorial biomarkers appeared around 2010, and numerous authors simply use the term in the following sense: If---say---the high concentration of all the molecules $A$, $B$ and $C$ characterizes a certain condition $X$ well (and the high concentration of any subset of the set $\{A,B,C\}$ would not), then they say that $\{A,B,C\}$ is a combinatorial biomarker of the condition $X$ \citep{Wu2012}.  In \citep{O'Bryant2011}, by applying  proteomics assays, a 30-protein set was identified as a combinatorial biomarker of AD. 

We intend to discover more involved combinatorial biomarkers that may contain clinical laboratory data and psychiatric test data, and we count not only on positive findings (i.e., high concentration or appearance of a certain value), but also the lack thereof (i.e., normal or low concentration). We start with frequent item set analysis, then apply association rule mining \citep{PDM}. We apply a new methodology that discovers complex combinatorial biomarkers only if these complex biomarkers have stronger implications than the simpler biomarkers. 

Therefore, our program SCARF will not produce artificially complex biomarkers: the more complex the new biomarker, the more valid the new implication.

\subsection{Association Rule Mining} Our research group was among the first applying association rule mining in molecular biology \citep{Ivan2007}. Recently, association rule mining has been gaining applications in drug discovery \citep{Galustian2010}, in the design of clinical trials \citep{Epstein2009}, and most recently, also in image analysis in Alzheimer's research \citep{Chaves2011}.

Association rule mining is a field of data mining \citep{PDM} developed by marketing experts for discovering implication-like rules in uncovering customer behavior \citep{Assoc}, without {\em a priori} assumptions on this behaviour. We intended to apply this method for laboratory and  cognitive test data from the CAMD database \citep{Romero2009}.

We analyzed how the presence/absence/severity of cognitive impairment could be detected from combinations of known biomarkers, demographic information and measurements of vital signs. As an example, consider this expression:

\begin{equation}
sodium=high \ \& \ (protein=high \ \hbox{or}\  age \ge 60) \rightarrow mmse\_total \le 15
\end{equation}

Here \& stands for logical AND. This rule states that if blood sodium is high, AND urine protein is high OR age is at least 60, then the total MMSE (Mini Mental State Examination) score will be at most 15 out of 30. Let us call the left-hand side of the expression (abbreviated by LHS) a combinatorial marker of the right-hand side (abbreviated by RHS). Thus the statement above can be reformulated as follows: high serum sodium combined with either high urine protein or age of at least 60 is a marker of a total MMSE score less than or equal to 15.

An expression consists of {\em elementary clauses} combined by logical operators. These elementary clauses may include equalities and inequalities. By substituting all elementary clauses with some wildcard, we can obtain the {\em pattern} of an expression. For example, the expression above is of the following pattern:

\begin{equation}
\square \ \&\  (\square \ \hbox{or}\  \square) \rightarrow \square
\end{equation}

During our analysis we started with a given pattern like the one above. Then we considered all the possible logical expressions according to this pattern, and assigned numerical values to them that indicated the reliability and validity of the logical rules. Then we filtered and sorted the vast amount of possible rules according to these numerical criteria, and selected the best ones. We changed a simpler rule to a more complex rule only if the more complex rule had higher reliability/validity than the simpler rule (see the next section for the exact definitions).

\section{Materials and Methods}

Our data source, which will be referred to as CAMD from now on \citep{Romero2009}, was provided by the Coalition Against Major Diseases, and consisted of the placebo arm of several drug trials. Over 6000 subjects participated in these trials including demented and not demented people of various age and sex (see Table \ref{egy}) for basic statistics). Standard laboratory data that have been collected for the subjects included about 300 different values in blood or urine altogether. These values were generally measured multiple times per subject (on different visit days), though each person was tested for only about 30 different values. The cognitive and psychological status of the subjects was measured at different times by standardized questionnaires ADAS-COG, ADCS-ADL, MMSE, NPI and SIB. In addition, some genetic tests have been performed, e.g., ApoE and MTHFR genotypes were recorded. Vital sign measurements (BP, pulse rate, respiratory rate and body temperature) have also been taken. Results concerning this dataset will be described in greater detail below.

We transformed this large dataset into a conveniently processable form. The CAMD database contained several rows describing one person and these were scattered between multiple data tables. So we collected the essential data from CAMD into one single table: this simplified table contained only one row for each subject.

If a subject was tested on different visit days, then we took the average of these test results. The resulting main table for CAMD consisted of around 170 columns (record fields) and 6000 rows (entries).

Our main method of processing the resulting table was association rule mining. First, we took a given pattern like $\square \ \&\  (\square \ \hbox{or}\  \square) \rightarrow \square$. Notice that the LHS (Left Hand Side) is in conjunctive normal form here (multiple OR clauses ANDed together). This pattern can be encoded as ``1 2'', as the first OR clause has one sub-clause and the second one has two. This pattern matches all statements of the following kind: ``if property A is present and property B or property C is present, then property D is present''.

Since we are interested in implication-like association rules that indicate factors implying normal or demented mental state, we made restrictions on which data columns can occur on the LHS (Left Hand Side) and the RHS (Right Hand Side). Laboratory data and sex were allowed on the LHS, and columns directly indicating mental status on the RHS. Then we gave numerical constraints on the ``goodness'' of a rule---thus introducing an ordering on the rules. Finally we tried to fill in all the void boxes in all possible ways to find the best rules.

If done without any optimization, this process would have yielded a vast amount of different rules that would have needed to be evaluated ``by hand''. Even just enumerating all the possible matches to this pattern would have required enormous computational resources. Consequently, we needed to make the computation feasible: we used a {\em branch-and-bound} approach similar to the Apriori Algorithm \citep{PDM}: if certain values for the first two boxes made a rule fail our constraints---regardless of what would be written in the third box---, then we threw out the rule and did not bother checking all the possible values for the third box. (A good analogue would be cutting a tree in a clever way: one does not bother removing all the little twigs one by one, but rather cuts the trunk.) This technique saved us considerable computational time, and made possible this study.

The association rule mining was done with our own program written in the C++ programming language, named SCARF (Simple Combinatorial Association Rule Finder). We calculated various standard numerical values for all association rules, which would indicate their validity. First, we defined the \em universe \em of a rule: this is the set of the database rows where all columns present in the rule have a known value. As we mentioned before, not all subjects were tested for everything, so our database contained a large amount of N/A entries. For testing the validity of a rule, only those rows could be taken into account, where there is no N/A written to any of the columns participating in the rule.

For evaluating the validity of a rule, we continued to work with only its universe and temporarily discarded all other rows in the database. Next, we calculated the \em LHS support\em, \em RHS support \em and \em support \em of a rule. The \em LHS support \em is the number of the rows where the LHS is true, the \em RHS support \em is the number of the rows where the RHS is true, and the \em support \em is the number of the rows where both the LHS and the RHS are true.

Then, we calculated the \em confidence\em, \em lift\em, \em leverage \em and \em $\chi^2$-statistic \em for a rule. The \em confidence \em is defined as the conditional probability of the RHS, assuming that the LHS is true. If one has high serum sodium combined with high urine protein or age at least 60 in our example, then confidence describes the chance of having a low MMSE score. The \em lift \em shows how many times the presence of the LHS increases the probability of RHS. Generally it indicates how big a risk factor the LHS is---though it is not certain that the LHS \em causes \em the RHS, as they both may be only consequences of some background phenomenon \citep{PDM}.

The \em leverage \em is the difference between the observed probability of both the LHS and RHS being true, and the estimated probability we get by assuming that the LHS and RHS are independent events. It indicates the level of dependency between the LHS and the RHS in a way. Finally, the \em $\chi^2$-statistic \em is a well-known measure of the estimated dependence of the indicator variables of the LHS and RHS. The \em p-value \em output by SCARF comes from this $\chi^2$ test.

The \em E-value \em (also calculated by SCARF) equals to the p-value multiplied by the total number of possible rules. The E-value is a more useful measure of randomness, since if we examine many rules, there is a high probability that the p-value will be small enough, while the E-value is insensitive for this kind of artifact.

The following table formalizes some of the above definitions. Here $P$ denotes the probability measure, and $P(A|B)$ denotes the conditional probability of event $A$ on condition $B$:

\begin{eqnarray*}
\mathrm{Confidence} & = & P(RHS|LHS) \\ \\
\mathrm{Lift} & = & \frac{P(RHS|LHS)}{P(RHS)} \\ \\
\mathrm{Leverage} & = & P(RHS \wedge LHS) - P(RHS)P(LHS)
\end{eqnarray*}


 For the CAMD database the acceptable values were set as follows: $\mathrm{universe} \geq 500$, $\mathrm{support} \geq 50$, $\mathrm{confidence} \geq 0.5$, $\mathrm{lift} \geq 1.2$, $\mathrm{p-value} \leq 0.05$.  In particular, we recorded rules on data that were measured on at least 500 subjects. We defined the \em goodness \em of a rule to be equal to its lift.
 
 Therefore we listed association rules of lift at least 1.2, i.e., only those rules were listed where the LHS increased the probability of RHS with at least 20\%.

One of the most significant novelties in our approach was filtering out those rules which were too complicated. The SCARF program threw out elementary clauses from the LHS as long as the overall goodness (i.e. the lift) of the rule did not decrease by more than $2\%$. Then it deleted the whole rule if its numerical values dropped below our constraints during the simplification process. In other words, we sacrificed some of the lift for simplicity, to avoid overfitting.

Having listed the best rules, we also tried to determine whether the elementary clauses (like $lb\_ast=h$, $lb\_folate=l$, etc.) have positive or negative effect on mental state. Therefore we counted their appearances on LHS, and classified these occurrences by the nature of the RHS: does it indicate normal cognition or rather dementia? We counted how many times an elementary clause occurred on the LHS of a rule when the RHS indicated a positive mental state, and how many times it occurred in rules where the RHS showed a negative state. Thus, in addition to mining rules whose LHS could probably serve as good combinatorial risk factor of dementia, we estimated the contribution of the \em individual \em clauses, for example ``protein=$high$'' to the onset of cognitive impairment.

For an elementary clause, \em Positive score \em was the number of rules with positive RHS, and \em Negative score \em was the number of rules with negative RHS. Then we compared \em Positive score \em with \em Negative score \em: by subtracting the negative score from the positive score we got a value called simply the {\em score} of the clause. Those elementary clauses whose score was positive were called {\em positive} clauses, and similarly, those where the score was negative were called {\em negative} clauses.

To summarize our method: we searched for combinatorial biomarkers using a branch-and-bound algorithm for association rule mining; then made statistical analysis regarding elementary clauses.

\section{Results}

The program outputs 725 rules from the CAMD database. Selected rules, ordered by lift (i.e. ``goodness'') decreasing are listed in Table \ref{ketto}. The whole set of rules is presented as Table S1 of the online supporting material.

On the LHS, clauses concerning biomarkers end in ``=l'', ``=h'', ``=n'', or combinations of these. Here $l$ means low, $h$ means high and $n$ means normal. If there are multiple letters (such as $nh$), then the corresponding equality states that the value is either high or normal. In other words, single letters correspond to a value category, while multiple letters mean the union of these categories.

For example, the second rule in Table \ref{ketto} was that of the second best lift. It can be interpreted in the following way: It is likely that if serum sodium level is elevated, and serum glucose level is either too low or normal, then the total MMSE score will be less than 15. Note that it is true for all rules of ours that there is not necessarily a causal relation between the LHS and RHS, as both the LHS and RHS can be consequences of an unknown process in the background.

The third rule states that ``if serum sodium level is elevated, and calcium level is either low or normal, then MMSE orientation subscore will be at most 2''. 
The seventh rule in Table \ref{ketto} states that ``if serum sodium level is elevated, and body temperature is too low, then total MMSE score will be less than 15''.

From these selected rules we can conclude that elevated sodium combined with various other factors (not too high glucose, not too high calcium, low temperature) might be a good indicator (or even the cause) of mental decline.

Elementary clauses with the greatest positive effect on normal cognition are listed in Table \ref{ot}.

Elementary clauses with the greatest negative effect on normal cognition are listed Table \ref{hat}.

\section{Discussion}

Among the 725 rules identified, 513 had lift values exceeding 2.00. Most of the rules exceeding even the 3.00 lift value had one thing in common: the LHS contained the premise $lb\_sodium=h$.

\subsection{Liver function} 
The rules found suggest that having high serum levels of AST (aspartate aminotransferase), as well as having low or high serum levels of ALT (alanine aminotransferase) may predispose to an impaired cognition characterized by low mini mental state examination (MMSE) scores. It should be noted that low ALT was much more rare in the CAMD database than high ALT, so the negative effect should be attributed mainly to high ALT. However, serum ALP (alkaline phosphatase) levels seem to have a controversial effect on mental status.

AST, ALT and ALP levels derive from the liver. Elevated ALP might indicate bile duct obstruction. AST or ALT may elevate in a number of cases of liver injury or damage, spreading from acute or chronic viral infections to alcohol induced or non-alcoholic steatohepatitis. It is interesting to note that elevated serum levels of AST (more than those of ALT) have been associated with impaired mental status. Although mild elevations in serum levels of AST and ALT are nonspecific to the etiology of liver injury, certain alteration patterns in these parameters may reflect the nature of the hepatic disease. For instance, the value of the AST/ALT ratio---also known as the De Ritis ratio---is approximately 0.8 in normal subjects, a ratio exceeding 2.00 being suggestive to alcoholic hepatitis.

Therefore we scanned the subjects with high AST values for higher than 2 AST/ALT ratio: we have only found 10 subjects satisfying both conditions. In addition, only 2 rules had AST/ALT on the left-hand side. Consequently, we may assume that high serum AST in the study subjects is not typically accompanied with high De Ritis ratio (i.e. probable alcoholic hepatitis).

The association of impaired liver function with mental decline can be illuminated in two perspectives. On one hand, impaired liver function might be insufficient to prevent the brain from the effects of certain neurotoxins, e.g. ammonia. This happens in the case of hepatic encephalopathy (HE), when severe liver damage resulting in acute liver insufficiency cannot detoxificate ammonia and other neurotoxins. On the other hand, the association of elevated AST/ALT ratio with impaired mental status proposes that another obscure element (e.g. chronic alcohol consumption) might be the factor responsible for both cognitive and metabolic damages.

Our results raise the possibility of a pathogenetic linkage between liver function and mental status in patients with AD.  Such linkage has also been proposed by other studies \citep{Sutcliffe2011,Astarita2010}. One study concludes that peripheral reduction of $\beta$-amyloid is sufficient to reduce brain $\beta$-amyloid and proposes that $\beta$-amyloids, which are of major pathogenic importance in AD may originate from the liver \citep{Sutcliffe2011}. Another research found that deficient liver production of docosahexaenoic acid (a neuroprotective fatty acid) correlates with impaired cognitive status in AD patients \citep{Astarita2010}.

To rule out the possibility when the elevated AST level is due to some medications taken, we compiled a detailed Table\_S3 (in the supporting on-line material) containing the number of subjects taking certain drugs, and the number of drug-takers with high AST. The data shows that, for example, 1929 subjects took Donepezil, while among the Donepezil-takers, only 415 have had high AST levels. 

\subsection{Serum sodium} A great number of rules (224) have high sodium on the left hand side, all of which have impaired cognition on the right hand side.
Net water loss is responsible for the majority of cases of hypernatremia \citep{Adrogue}. A recent publication, examining the causes and comorbidities in patients older than 65 years, has found that the most common cause of community-acquired hypernatremia is dehydration due to reduced oral intake \citep{Turgutalp2012}. More interestingly, they found that the most common comorbidity in this patient group was AD, present in 31.4\% of patients with hypernatremia \citep{Turgutalp2012}. Hydration status has a significant impact on the volume of grey and white matter in the brain and on the quantity of cerebrospinal fluid as a hallmark of ventricular enlargement \citep{Streitbuerger2012}. The pattern of shrinkage in white matter volume and increase of the ventricular system due to dehydration is consistent with the structural brain changes observed during the progression of AD \citep{Streitbuerger2012}. In another study, patients with AD underwent bioelectrical impedance vector analysis to assess the body cell mass and hydration status related to AD \citep{Buffa2010}. Results demonstrated a tendency towards dehydration in patients with AD \citep{Buffa2010}. Although the association of dehydration and AD is supported by these publications, the specific pathogenic nature of this association remains obscure  \citep{Turgutalp2012, Streitbuerger2012, Buffa2010}.

\subsection{Vitamin B12} Our results were able to present the beneficial impact of high levels of vitamin B12, also known as cobalamin, on cognition. Along with folate, vitamin B12 has an important role in the maintenance of genome integrity \citep{Fenech2012}. Although previous publications found association of low serum levels of vitamin B12 and AD \citep{Malaguarnera2004,McCaddon2004}, a recent systemic review on vitamin B12 status and cognitive impairment fails to declare a clear association between vitamin B12 status and dementia \citep{OLeary2012}. However, this review also found that studies using newer and more specific biomarkers of vitamin B12 status such as methylmalonic acid and holotranscobalamin were able to draw an association between mental decline and poor vitamin B12 status \citep{OLeary2012}. 

Although clinically vitamin B12 deficiency may result in macrocytic anaemia, in the case of AD patients the occurrence of macrocytic anaemia is rare and the neurological and hematological features are unrelated \citep{McCaddon2004}. 

\subsection{Hematological parameters} 

Additional interesting rules were detected regarding hematological parameters. In particular, independently from each other, high values of mean corpuscular hemoglobin (MCH), low values of mean corpuscular hemoglobin concentration (MCHC), and low values of mean corpuscular volume (MCV) were also associated with high MMSE scores.
Although high values of MCH and low values of MCHC are present in the case of macrocyctic anaemia (with the addition of high levels of mean corpuscular volume, low levels of hemoglobin and hematocrit), such solely associations should not be discussed, as they may be coincidental. 

Among the rules with lift values exceeding 2.00, other parameters of hematological status (such as level of hemoglobin, red blood cell number, white blood cell number) were also present. Monocyte and eosinophil levels also appear on the left hand side of many rules with high lift. These premises appear in combinations with various other (mostly non-hematological) premises.

\subsection{Blood cholesterol and cognition}

The positive or negative effects of high cholesterol values to Alzheimer's disease and cognition is a controversial issue. Some studies (e.g., \citep{Helzner2009,Whitmer2005,Zambon2010}) show negative effects of high cholesterol value for cognition, while other studies (\citep{Reitz2004,Reitz2005,Mielke2005}) prove the positive effects for cognition.

Our data supports both conclusions in a sense. That is, low, low-normal and high cholesterol levels are all associated with impaired mental status, but with a different extent (scores -21, -13 and -42, respectively). See Table \ref{harom} for a selection of cholesterol-related rules from the larger Table S1 in the on-line supporting material.

It is worth to note that, by Table \ref{harom}, elevated, low or low-normal cholesterol levels do not necessarily mean a higher likelihood of impaired cognition by themselves, but only combined with high sodium.

A most recent study \citep{Pierrot} shows that the neuronal expression of amyloid precursor protein APP controls the cholesterol 24-hydroxylase mRNA levels and decreases cholesterol turnover; therefore in certain setups, the presence of amyloid precursor proteins imply lowered cholesterol levels.

\section{Conclusions} 

A 6000-patient, high-quality database was analyzed with original methods for biomarkers of dementia. 
We have found some novel and also some already well established relations connected to good or bad cognition in a 6000 patient database. The already established findings prove the validity of our datamining approach, and the new findings, related to MCH, ALP and AST levels prove its power. Some more controversial biomarkers, including cholesterol level, were also re-discovered, and we found that the high cholesterol levels seem to be beneficial only with combination with old age.


\vfill
\eject

\section*{Tables}

\begin{table}[ht]
\centering
  \caption{Basic statistics on the subjects of the CAMD data}
\small
 \begin{tabular}{lrlrlr}
 \toprule
{\textbf{Age distribution}} &   & {\textbf{Gender distribution}} &  & {\textbf{MMSE distribution}} &  \\
\midrule
          &       &       &              &       &         \\
    A: up to 65 years&  1093  & Female & 3315  & A: severe cog. impairment & 255 \\
    B: 66-75 years&  2070  & Male  & 2653   & B: moderate cog. impairment & 611 \\
    C: 76-85 years&  2408  &        &      & C: mild cog. impairment & 3224 \\
    D: more than 85&  397   &       &     & D: normal cognition & 1352 \\
\bottomrule
\end{tabular}

\label{egy}

\end{table}

\vfill			
\eject

\begin{table}[htbp]
\centering
  \caption{Several association rules of the highest lift. The lift value describes the multiplication factor, increasing the probability of the Right Hand Side (RHS) if the Left Hand Side is true. For example, our best rule (the first below) is saying that one can have the a bad result of a cognitive test with four times higher probability if one has high serum sodium and either low cholesterol or low or normal blood glucose level.}

\begin{verbatim}
(lb_sodium=h) & (lb_chol=l or lb_gluc=ln) ---> mm_ori=B
Universe: 2783, LHS support: 87, RHS support: 401, Support: 50
Confidence: 0.574713, Lift: 3.98859, Leverage: 0.0134618, p-value: 0, E-value: 0
3.98859

(lb_gluc=ln) & (lb_chol=l or lb_sodium=h) ---> mm_ori=B
Universe: 2783, LHS support: 105, RHS support: 401, Support: 57
Confidence: 0.542857, Lift: 3.76751, Leverage: 0.0150451, p-value: 0, E-value: 0
3.76751

(lb_sodium=h) & (lb_hct=l or lb_gluc=ln) ---> mm_ori=B
Universe: 2926, LHS support: 95, RHS support: 420, Support: 51
Confidence: 0.536842, Lift: 3.74, Leverage: 0.0127695, p-value: 0, E-value: 0
3.74

(lb_sodium=h) & (bpsys=ln or lb_gluc=ln) ---> mm_ori=B
Universe: 3091, LHS support: 102, RHS support: 425, Support: 52
Confidence: 0.509804, Lift: 3.70777, Leverage: 0.0122858, p-value: 0, E-value: 0
3.70777

(lb_gluc=ln) & (lb_creat=l or lb_sodium=h) ---> mm_ori=B
Universe: 3091, LHS support: 99, RHS support: 425, Support: 50
Confidence: 0.505051, Lift: 3.6732, Leverage: 0.0117722, p-value: 0, E-value: 0
3.6732

(lb_sodium=h) & (age=D or lb_gluc=ln) ---> mm_ori=B
Universe: 3091, LHS support: 101, RHS support: 425, Support: 51
Confidence: 0.50495, Lift: 3.67248, Leverage: 0.0120068, p-value: 0, E-value: 0
3.67248

(lb_gluc=ln) & (lb_ast=l or lb_sodium=h) ---> mm_ori=B
Universe: 3091, LHS support: 101, RHS support: 425, Support: 51
Confidence: 0.50495, Lift: 3.67248, Leverage: 0.0120068, p-value: 0, E-value: 0
3.67248
\end{verbatim}
\label{ketto}
\end{table}

\vfill
\eject

\begin{table}[htbp]
\centering
{\scriptsize
\caption{Some association rules involving serum cholesterol level}

\begin{verbatim}
(lb_sodium=h) & (lb_chol=l or lb_gluc=ln) ---> mm_ori=B
Universe: 2783, LHS support: 87, RHS support: 401, Support: 50
Confidence: 0.574713, Lift: 3.98859, Leverage: 0.0134618, p-value: 0, E-value: 0
3.98859

(lb_gluc=ln) & (lb_chol=l or lb_sodium=h) ---> mm_ori=B
Universe: 2783, LHS support: 105, RHS support: 401, Support: 57
Confidence: 0.542857, Lift: 3.76751, Leverage: 0.0150451, p-value: 0, E-value: 0
3.76751

(lb_sodium=h) & (lb_chol=ln or lb_gluc=ln) ---> mm_ori=B
Universe: 2783, LHS support: 106, RHS support: 401, Support: 55
Confidence: 0.518868, Lift: 3.60102, Leverage: 0.0142747, p-value: 0, E-value: 0
3.60102

(lb_chol=h) & (lb_cl=h or lb_sodium=h) ---> mm_ori=B
Universe: 1420, LHS support: 71, RHS support: 304, Support: 51
Confidence: 0.71831, Lift: 3.35526, Leverage: 0.0252113, p-value: 2.22045e-016, E-value: 1.88773e-007
3.35526

(lb_chol=h) & (lb_monole=l or lb_sodium=h) ---> mm_total=AB
Universe: 1364, LHS support: 73, RHS support: 325, Support: 58
Confidence: 0.794521, Lift: 3.33454, Leverage: 0.02977, p-value: 1.51101e-013, E-value: 0.00012846
3.33454

(lb_sodium=h) & (lb_monole=h or lb_chol=h) ---> mm_total=AB
Universe: 1364, LHS support: 66, RHS support: 325, Support: 51
Confidence: 0.772727, Lift: 3.24308, Leverage: 0.0258608, p-value: 5.9952e-015, E-value: 5.09687e-006
3.24308

(lb_chol=h) & (lb_hbsag=h or lb_sodium=h) ---> mm_attcal=B
Universe: 1164, LHS support: 67, RHS support: 312, Support: 50
Confidence: 0.746269, Lift: 2.78416, Leverage: 0.0275268, p-value: 6.2725e-011, E-value: 0.0533262
2.78416

(lb_sodium=h) & (lb_bun=h or lb_chol=h) ---> mm_attcal=B
Universe: 1387, LHS support: 61, RHS support: 429, Support: 52
Confidence: 0.852459, Lift: 2.75609, Leverage: 0.023888, p-value: 8.87168e-012, E-value: 0.00754232
2.75609

(lb_sodium=h) & (lb_ca=l or lb_chol=h) ---> mm_attcal=B
Universe: 1420, LHS support: 61, RHS support: 460, Support: 51
Confidence: 0.836066, Lift: 2.5809, Leverage: 0.0219996, p-value: 2.52266e-011, E-value: 0.0214466
2.5809

(lb_sodium=h) & (lb_cl=h or lb_chol=h) ---> mm_attcal=B
Universe: 1420, LHS support: 66, RHS support: 460, Support: 55
Confidence: 0.833333, Lift: 2.57246, Leverage: 0.0236759, p-value: 1.65421e-010, E-value: 0.140634
2.57246
\end{verbatim}
}
\label{harom}
\end{table}

\vfill
\eject

\begin{table}
{\tiny 
\centering
\caption{Legends for Table 2 and 3}

\begin{tabular*}{\textwidth}{@{\extracolsep{\fill}} | l | l | }
\hline
age & Subject age (A: $\leq$ 65 years, B: 66--75 years, C: 76--85 years, D: >85 years) \\ \hline
ast\_alt & De Ritis ratio \\ \hline
bpdia & Diastolic blood pressure \\ \hline
bpsys & Systolic blood pressure \\ \hline
lb\_alb & Serum albumine \\ \hline
lb\_alp & Serum alkaline phosphatase \\ \hline
lb\_alt & Serum alanine aminotransferase \\ \hline
lb\_ast & Serum aspartate aminotransferase \\ \hline
lb\_baso & Basophils, particle concentration \\ \hline
lb\_bili & Serum indirect bilirubin \\ \hline
lb\_bun & Blood Urea Nitrogen \\ \hline
lb\_ca & Serum calcium \\ \hline
lb\_chol & Serum cholesterol \\ \hline
lb\_ck & Serum creatine kinase \\ \hline
lb\_cl & Serum chlorine \\ \hline
lb\_creat & Serum creatinine \\ \hline
lb\_eos & Eosinophils, particle concentration \\ \hline
lb\_gluc & Serum glucose \\ \hline
lb\_hba1c & Hemoglobin A1C \\ \hline
lb\_hbsag & Hepatitis B virus surface antigen \\ \hline
lb\_hct & Hematocrit \\ \hline
lb\_hgb\_blood & Blood hemoglobin \\ \hline
lb\_k & Serum potassium \\ \hline
lb\_ketones & Ketones \\ \hline
lb\_ldh & Lactate dehydrogenase \\ \hline
lb\_lym & Lymphocytes, particle concentration \\ \hline
lb\_lymle & Lymphocytes/leukocytes ratio \\ \hline
lb\_mch & Mean Corpuscular Hemoglobin \\ \hline
lb\_mchc & Mean Corpuscular Hemoglobin Concentration \\ \hline
lb\_mcv & Mean Corpuscular Volume \\ \hline
lb\_mono & Monocytes, particle concentration \\ \hline
lb\_monole & Monocytes/leukocytes ratio \\ \hline
lb\_neut & Neutrophils, particle concentration \\ \hline
lb\_neutle & Neutrophils/leukocytes ratio \\ \hline
lb\_ph & pH \\ \hline
lb\_phos & Phosphate \\ \hline
lb\_plat & Platelets \\ \hline
lb\_prot & Total protein \\ \hline
lb\_rbc\_blood & Red blood count \\ \hline
lb\_sodium & Serum sodium \\ \hline
lb\_tsh & Thyrotropin \\ \hline
lb\_vitb12 & Serum B12 vitamin \\ \hline
lb\_wbc\_blood & White blood count \\ \hline
mm\_attcal & MMSE attention and calculation subscore (B: 0--1, C: 2, D: 3, E: 4--5) \\ \hline
mm\_lang & MMSE language subscore (B: 0--2, C: 3--4, D: 5--6, E: 7--9) \\ \hline
mm\_ori & MMSE orientation subscore (B: 0--2, C: 3--4, D: 5--7, E: 8--10) \\ \hline
mm\_recall & MMSE recall subscore (B: 0, C: 1, D: 2, E: 3) \\ \hline
mm\_total & MMSE total score (A: <10, B: 10--14, C: 15--23, D: $\geq$ 24) \\ \hline
pulse & Heart rate \\ \hline
resp & Respiratory rate \\ \hline
sex & Subject sex (F: female, M: male) \\ \hline
temper & Temperature \\ \hline
\end{tabular*}
\label{negy}
}
\end{table}

\begin{table}[htbp]
\centering
  \caption{Elementary clauses with the greatest positive effect on normal cognition}
\begin{tabular*}{\textwidth}{@{\extracolsep{\fill}} | r | l | }
\hline
lb\_vitb12=h & score: 67 \\ \hline
lb\_mch=h & score: 25 \\ \hline
lb\_mchc=l & score: 22 \\ \hline
lb\_k=h & score: 17 \\ \hline
sex=M & score: 10 \\ \hline
pulse=l & score: 9 \\ \hline
lb\_bun=l & score: 8 \\ \hline
age=AB & score: 4 \\ \hline
lb\_mono=nh & score: 3 \\ \hline
resp=ln & score: 3 \\ \hline
lb\_plat=ln & score: 2 \\ \hline
lb\_eos=nh & score: 2 \\ \hline
lb\_prot=nh & score: 2 \\ \hline
\end{tabular*}
\label{ot}
\end{table}

\begin{table}[htbp]
\centering
  \caption{Elementary clauses with the greatest negative effect on normal cognition}
\begin{tabular*}{\textwidth}{@{\extracolsep{\fill}} | r | l | }
\hline
temper=nh & score: -10 \\ \hline
lb\_wbc\_blood=h & score: -10 \\ \hline
age=BCD & score: -10 \\ \hline
lb\_prot=h & score: -12 \\ \hline
lb\_gluc=h & score: -12 \\ \hline
pulse=h & score: -12 \\ \hline
lb\_ck=h & score: -12 \\ \hline
lb\_hct=nh & score: -12 \\ \hline
lb\_k=ln & score: -12 \\ \hline
lb\_alp=h & score: -12 \\ \hline
lb\_chol=ln & score: -13 \\ \hline
lb\_ph=h & score: -13 \\ \hline
lb\_hct=l & score: -13 \\ \hline
lb\_alt=h & score: -13 \\ \hline
age=A & score: -14 \\ \hline
bpsys=ln & score: -14 \\ \hline
lb\_creat=ln & score: -14 \\ \hline
lb\_creat=h & score: -16 \\ \hline
temper=l & score: -17 \\ \hline
lb\_alp=ln & score: -18 \\ \hline
lb\_bun=ln & score: -18 \\ \hline
lb\_alt=l & score: -19 \\ \hline
lb\_wbc\_blood=l & score: -20 \\ \hline
lb\_chol=l & score: -21 \\ \hline
pulse=nh & score: -21 \\ \hline
lb\_prot=ln & score: -22 \\ \hline
lb\_bun=h & score: -22 \\ \hline
lb\_plat=h & score: -26 \\ \hline
lb\_gluc=ln & score: -27 \\ \hline
bpdia=ln & score: -28 \\ \hline
age=CD & score: -32 \\ \hline
lb\_chol=h & score: -42 \\ \hline
lb\_ast=h & score: -43 \\ \hline
lb\_ca=l & score: -50 \\ \hline
sex=F & score: -57 \\ \hline
age=D & score: -99 \\ \hline
lb\_cl=h & score: -173 \\ \hline
lb\_sodium=h & score: -224 \\ \hline
\end{tabular*}
\label{hat}
\end{table}

\end{document}